\documentclass[conference]{IEEEtran}
\IEEEoverridecommandlockouts
\usepackage{cite}
\usepackage{amsmath,amssymb,amsfonts}
\usepackage{algorithm}
\usepackage{algpseudocode}
\usepackage{graphicx}
\usepackage{textcomp}
\usepackage{xcolor}
\def\BibTeX{{\rm B\kern-.05em{\sc i\kern-.025em b}\kern-.08em
    T\kern-.1667em\lower.7ex\hbox{E}\kern-.125emX}}
\begin{document}

% \title{LGCP: Clustered Hierarchical Collaborative Perception for Global Awareness\\
\title{Efficient Local-to-Global Collaborative Perception via Joint Communication and Computation Optimization\\

% \thanks{This work was supported by the National Natural Science Foundation of China with Grant  No. 62231015.}
}

\author{\IEEEauthorblockN{Hui Zhang, Yuquan Yang, Zechuan Gong, Xiaohua Xu}
\IEEEauthorblockA{\textit{School of Computer Science and Technology} \\
\textit{University of Science and Technology of China}\\
Hefei, China \\
% \{fzhh,xiaohuaxu\}@ustc.edu.cn, \{yuquany,gongzechuan\}@mail.ustc.edu.cn
}
\and
% \IEEEauthorblockN{Yuquan Yang}
% \IEEEauthorblockA{\textit{School of Computer Science and Technology} \\
% \textit{University of Science and Technology of China}\\
% Hefei, China \\
% yuquany@mail.ustc.edu.cn}
% \and
% \IEEEauthorblockN{Zechuan Gong}
% \IEEEauthorblockA{\textit{School of Computer Science and Technology} \\
% \textit{University of Science and Technology of China}\\
% Hefei, China \\
% gongzechuan@mail.ustc.edu.cn}
% \and
% \IEEEauthorblockN{Xiaohua Xu}
% \IEEEauthorblockA{\textit{School of Computer Science and Technology} \\
% \textit{University of Science and Technology of China}\\
% Hefei, China \\
% xiaohuaxu@ustc.edu.cn}
% \and
\IEEEauthorblockN{Dan Keun Sung}
\IEEEauthorblockA{\textit{School of Electrical Engineering} \\
\textit{Korea Advanced Institute of Science and Technology}\\
Daejeon, Korea \\
% dksung@kaist.ac.kr
}
}

\maketitle

\begin{abstract}
Autonomous driving relies on accurate perception to ensure safe driving.
Collaborative perception improves accuracy by mitigating the sensing limitations of individual vehicles, such as limited perception range and occlusion-induced blind spots.
However, collaborative perception often suffers from high communication overhead due to redundant data transmission, as well as increasing computation latency caused by excessive load with growing connected and autonomous vehicles (CAVs) participation. 
To address these challenges, we propose a novel local-to-global collaborative perception framework (LGCP) to achieve collaboration in a communication- and computation-efficient manner. 
The road of interest is partitioned into non-overlapping areas, each of which is assigned a dedicated CAV group to perform localized perception.
A designated leader in each group collects and fuses perception data from its members, and uploads the perception result to the roadside unit (RSU), establishing a link between local perception and global awareness.
The RSU aggregates perception results from all groups and broadcasts a global view to all CAVs.
LGCP employs a centralized scheduling strategy via the RSU, which assigns CAV groups to each area, schedules their transmissions, aggregates area-level local perception results, and propagates the global view to all CAVs.
Experimental results demonstrate that the proposed LGCP framework achieves an average 44× reduction in the amount of data transmission, while maintaining or even improving the overall collaborative performance.

\end{abstract}

\begin{IEEEkeywords}
Connected and autonomous vehicles, Local-to-global, Collaborative perception.
\end{IEEEkeywords}

\section{Introduction}
\begin{figure}    
	\centering
    \includegraphics[width=0.9\linewidth]{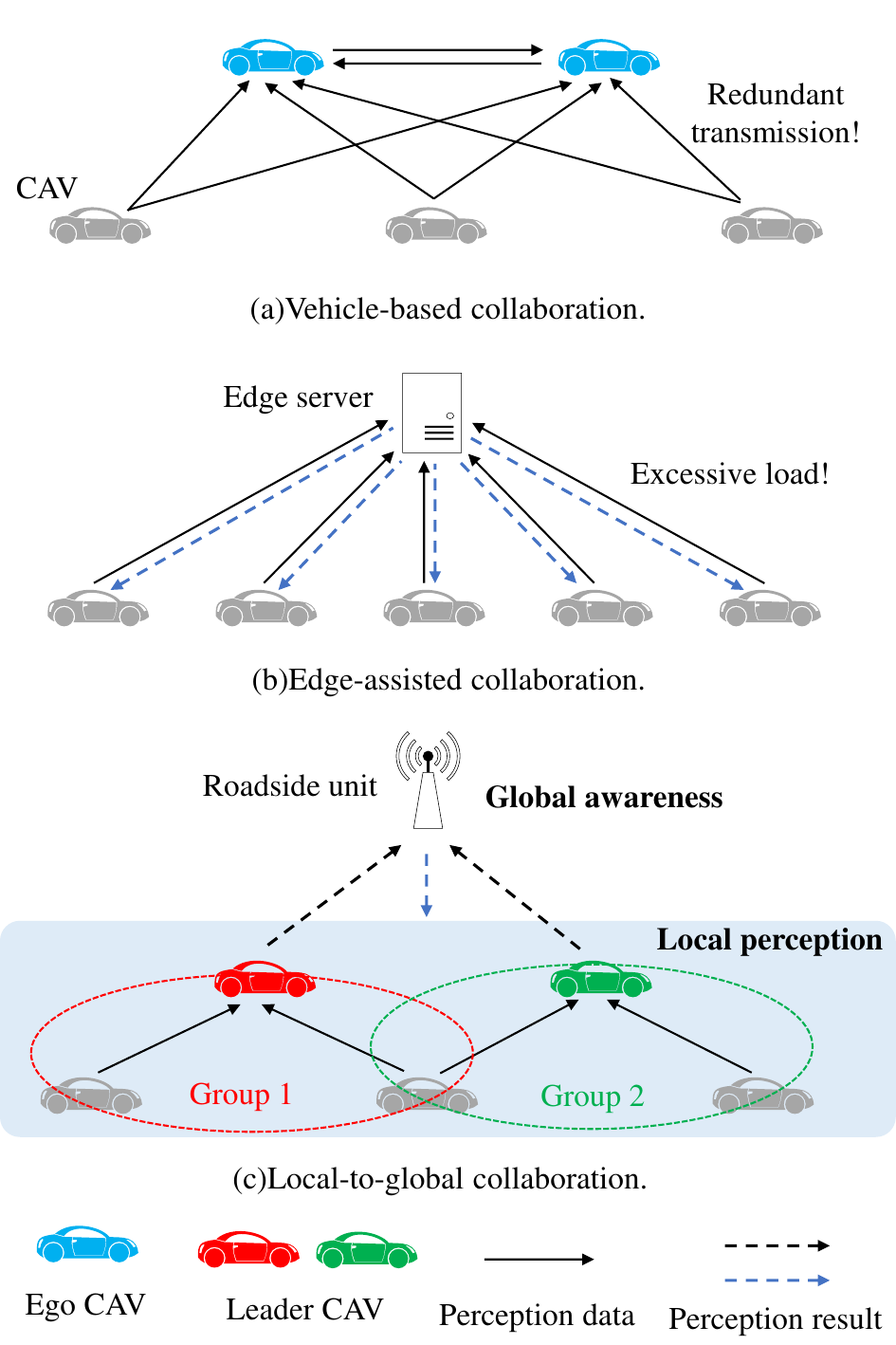}
    \caption{Different collaboration perspectives. }
    \label{fig:intro}%文中引用该图片代号
\end{figure}

Autonomous vehicles have the potential to reduce human driving errors, and enhance road safety, thereby enabling more intelligent traffic management and mobility services. 
It is crucial to perceive and comprehend complex driving environments for effective operation of autonomous vehicles.
Advances in localization and perception enable even single-vehicle systems with high-precision GPS, cameras, and LiDAR to achieve notable performance in tasks like object detection and tracking\cite{2015ITITS-chavez-garcia-MultipleSensorFusion,2023ITPAMI-zheng-EffectiveMotionCentricParadigm,2023SR-liu-RealTimeObject}. 
However, a number of limitations imposed by sensor characteristics, lighting conditions, obstacle occlusion and other factors may restrict the perception range of single-vehicle systems, leading to blind spots in which autonomous vehicles may not detect the approaching dangers in real-time conditions\cite{2021PICCVPR-yuan-RobustInstanceSegmentation,2021ITCSVT-yuan-TemporalChannelTransformer3D}. 
To further improve the perception capability of autonomous vehicles, collaborative perception enables multiple connected and autonomous vehicles (CAVs) to engage in vehicle-to-everything~(V2X) communications to obtain broader environmental data\cite{2020CV21ECGUA22PPI1-wang-V2VNetVehicletoVehicleCommunication}. 
Collaborative perception may overcome some inherent limitations of a single vehicle perception system, such as long-distance perception and occlusion, by aggregating the perception data from multiple CAVs, which can effectively extend the perception range, improve the perception accuracy, enhance the CAV's situational awareness and responsiveness, and improve the reliability and safety of CAVs in diverse traffic environments\cite{2022IRAL-yuan-KeypointsBasedDeepFeature,2025pr-chu-occlusionguidedmultimodalfusion, 2025pr-liu-sparsecommefficientsparse}. 
Therefore, collaborative perception has been recognized as a pivotal technology for the advancement of autonomous driving.

In purely vehicle-based collaboration~\cite{2021ANIPS-li-LearningDistilledCollaboration,2022APA-xu-CoBEVTCooperativeBirds,2022ANIPS-hu-Where2commCommunicationEfficientCollaborative,20232IICRAI-lu-RobustCollaborative3D}, the ego CAVs aggregate perception data from other neighboring CAVs, and then fuses the received perception data, as shown in Fig.\ref{fig:intro}(a).
Though the vehicle-based collaboration paradigm benefits from other neighboring CAVs, it still faces some inherent limitations. 
One limitation is redundant transmission.
When each ego CAV individually requests perception data from other surrounding CAVs, a large amount of redundant perception data may be exchanged among CAVs.
The other limitation is its limited perception range.
Due to the unreliable wireless communication among CAVs, such as link degradation over long distances, the ego CAV may fail to receive perception data from distant CAVs.
In particular, when traffic density is high, wireless signals between CAVs may be obstructed by other vehicles \cite{2021ITVT-lv-BlockageAvoidanceBased,2024itvt-zhang-efficientvehicularcollaborative}, which limits CAVs to local perception.

A roadside edge server can engage in the collaborative perception process, as shown in Fig.\ref{fig:intro}(b).
In the edge-assisted collaboration paradigm~\cite{2021ii2ccc-liu-edgesharingedgeassisted,2021P2AICMCN-zhang-EMPEdgeassistedMultivehicle, 2024P2ACENSS-zhu-BoostingCollaborativeVehicular, 2023IJSAC-luo-EdgeCooperNetworkAwareCooperative}, the edge server aggregates perception data from adjacent CAVs, fuses the received data, and propagates the perception results back to all CAVs.
Since each CAV only needs to transmit perception data to the edge server, redundant data transmissions are significantly reduced.
Due to higher-positioned antennas, the edge-assisted collaboration may achieve a wider perception range, compared with the vehicle-based collaboration, enabling global awareness.
However, the centralized aggregation and processing of perception data from all adjacent CAVs at the edge server may lead to transmission and computation latency that grows linearly with the number of participating CAVs.
Correspondingly, when the amount of traffic is large, the edge-assisted collaboration may suffer from long transmission and computation latency~\cite{20232ISECS-liu-AdaMapHighscalableRealtime, 2024I2ICC-zhang-EfficientFusionTreeBasedPerception}, resulting in excessive load.

To address the above problems, we consider the collaboration paradigm from a local-to-global perspective, which combines the strengths of distributed and centralized strategies.
For an area within a road of interest (RoI), multiple surrounding CAVs exhibit varying levels of perception accuracy for that area.
Selecting an appropriate group and designating a leader from these CAVs to perform the area-specific collaboration can help achieve a good balance between accuracy and the overhead of transmission and computation.
% The RoI can be considered as a set of non-overlapping areas, formed by first partitioning it into grid cells, assigning a CAV group to each cell, and merging cells assigned the same group.
The RoI can be considered as a set of non-overlapping areas.
By adopting a divide-and-conquer strategy, local perception result within each area is processed by its group leader. 
Aggregating perception results of all areas can yield a global awareness of the RoI, which is then broadcast back to all CAVs.
This local-to-global design reduces transmission overhead through centralized assignment that determines what data to share and with whom among CAVs, while computation is distributed across CAVs to lower latency, as shown in Fig.\ref{fig:intro}(c).
% For an area within a road of interest (RoI), objects within the area can be perceived by multiple surrounding CAVs with varying levels of perception accuracy.
% Selecting an appropriate cluster from these CAVs can help achieve a good balance between collaboration accuracy and the overhead of transmission and computation.
% The RoI can be considered as a set of non-overlapping areas, each of which is assigned a dedicated CAV cluster to finish collaborative perception within its area.
% Aggregating perception results of all areas can yield a global view of all the observed objects within the RoI, which is then broadcast to all CAVs.
% This hierarchical design enables selective sharing of perception data among CAVs within the same area, reducing transmission cost, while distributed processing leverages collective resources to lower computational latency, as shown in Fig.\ref{fig:intro}(c).
% One problem is how to partition the RoI. 
% To enable dynamic area formation, we first divide the RoI into grid cells, assign a CAV group to each cell, and merge cells assigned the same group.
% However, how to assign a satisfactory CAV group to each cell.
The key challenge lies in assigning a satisfactory CAV group to each area.
While assigning more CAVs generally improves perceptual quality, it may also increase transmission and computation overhead, and introduce communication issues such as link interference and degradation.
To solve this problem, we introduce a confidence metric to evaluate the perceptual quality of a CAV for a specific area, and formulate the assignment problem as an optimization problem that maximizes the total confidence under real-time constraints.

Based on the above ideas, we propose a \textbf{L}ocal-to-\textbf{G}lobal \textbf{C}ollaborative \textbf{P}erception (LGCP) framework. 
LGCP adopts a centralized scheduling policy to assign different CAV groups for different areas.
A roadside unit (RSU) partitions the RoI into non-overlapping areas and broadcasts an initiation message to start a global collaboration.
Each CAV engages in the collaboration by sending its basic information such as location, direction and its confidence values for perceived areas to the RSU.
Based on the received information, the RSU assigns the perception task of each area to a specific CAV group. 
A designated leader in each group collects and fuses perception data from its members, and uploads the perception result of the corresponding area to the RSU, establishing a link between local perception and global awareness.
The RSU aggregates perception results from all groups and broadcasts a global awareness result to all CAVs engaged in the collaboration. 
The contributions of this paper are as follows:
\begin{itemize}
    \item We propose a novel local-to-global collaborative perception framework for the perception tasks of autonomous driving systems.
    The proposed framework can enhance collaborative perception efficiency in a communication- and computation-efficient manner by partitioning the RoI into non-overlapping areas and aggregating the perception results of all areas. 

    \item We consider transmission cost and computation latency as collaborative perception performance indices, introduce a confidence metric to evaluate the perceptual quality of a CAV for a specific area, and formulate the trade-off between perception performance and resource overhead as an optimization problem. 
    We propose a greedy algorithm for assigning CAV groups, and a transmission scheduling algorithm for the transmission of these CAVs.
    
    \item The LGCP framework adopts existing collaborative perception models for the perception tasks of areas.
    Experimental results demonstrate that the proposed LGCP framework achieves an average 44× reduction in the amount of data transmission, while maintaining or even improving the overall collaborative performance.

\end{itemize}

\section{Related Work}
% \subsection{Vehicle-based Mechanisms}
Depending on the stage of information transmission and fusion, collaborative perception modes can mainly be categorized into raw-data-based early collaboration, feature-based intermediate collaboration, and object-based late collaboration.
Early collaboration aggregates and aligns raw sensor data from different CAVs, and then detects objects based on the aligned perception data.
Chen et al. \cite{20192I3ICDCSI-chen-CooperCooperativePerception} first proposed a sparse-point-clouds-based object detection method, demonstrating the feasibility of early collaboration.
Although early collaboration may ideally achieve optimal perception accuracy, it incurs substantial transmission overhead due to the large volume of raw sensor data \cite{20222ICRAI-xu-OPV2VOpenBenchmark}.
In late collaboration, the perception results of objects are usually represented as a set of boxes.
Compared with early collaboration, late collaboration can achieve much higher data aggregation efficiency due to lower transmission cost of perception object results.
Xu et al. \cite{20232IICRAI-xu-ModelAgnosticMultiAgentPerception} proposed an offline-trained confidence calibrator to remove the effects of prediction confidence bias.
However, the inherent weakness in individual perception may mislead collaborative efforts, resulting in unsatisfactory performance.

Considering the limitations of early and late collaboration, more researchers have focused on intermediate collaboration.
Xu et al. \cite{2022APA-xu-CoBEVTCooperativeBirds} proposed a fused axial attention module to efficiently aggregate multi-view and multi-agent features by capturing local and global spatial interactions within a transformer architecture.
Hu et al. \cite{2022ANIPS-hu-Where2commCommunicationEfficientCollaborative} proposed using a spatial confidence map to enable agents to share only sparse but perceptually critical features, contributing to the trade-off between communication efficiency and perception performance.
Zhang et al. \cite{2024itc-zhang-collaborativemultimodalfusion} proposed modality-aware fusion to better integrate sensory inputs, using multi-view fused point clouds as dense supervision for self-supervised depth prediction.
Some other researchers proposed jointly utilizing multiple kinds of perception data to improve perception performance.
% Li et al. \cite{2021ANIPS-li-LearningDistilledCollaboration} proposed a knowledge extraction framework that uses early collaboration to guide intermediate model training, balancing performance and communication cost.
Lu et al. \cite{20232IICRAI-lu-RobustCollaborative3D} proposed agent-object pose graph modeling to enhance pose consistency, along with multiple collaboration strategies for aggregating intermediate features at multiple spatial resolutions.
The above vehicle-based approaches usually focus on the performance of the proposed collaboration models in terms of perception accuracy and the amount of transmission data, almost ignoring the impact of wireless communication characteristics on the collaborative process.
Although Ren et al. \cite{2024itiv-ren-interruptionawarecooperativeperceptiona} proposed a communication adaptive spatial-temporal prediction model that can mitigate the impact of missing information by leveraging historical information, it cannot fundamentally address the inherent vulnerability of the vehicle-based paradigm to communication challenges such as link interference and degradation.

% \subsection{Edge-assisted Mechanisms}
Some researchers have considered edge-assisted collaboration among CAVs.
Liu et al.\cite{2021ii2ccc-liu-edgesharingedgeassisted} proposed uploading raw sensor data from multiple CAVs to an edge server to build a global perception view.
To reduce the transmission cost, Zhang et al.\cite{2021P2AICMCN-zhang-EMPEdgeassistedMultivehicle} proposed transmitting partitioned raw sensor data of CAVs, and Luo et al.\cite{2023IJSAC-luo-EdgeCooperNetworkAwareCooperative} proposed uploading complementarity-enhanced and redundancy-minimized raw sensor data of CAVs. 
Hou et al.\cite{2025ITITS-hou-EnhancingCooperativeLiDARbaseda} proposed selecting CAVs capable of achieving high collaborative perception performance for data upload.
Considering the impact of unreliable wireless links between CAVs and the edge, Zhu et al.\cite{2024P2ACENSS-zhu-BoostingCollaborativeVehicular} proposed using vehicles as relays in a multi-hop manner to facilitate raw sensor data upload to the edge.
However, multi-hop transmission introduces additional overhead, which may make single-hop communication more preferable in practice.
Considering the high transmission cost of raw sensor data, Lin et al.\cite{2024ITWC-lin-EdgeassistedCollaborativePerception} proposed to upload the data features of CAVs to the edge and proposed an edge-assisted reinforcement learning-based collaborative perception scheme, and Lin et al.\cite{2024ITMC-lin-EdgeAssistedLowlatency} proposed to upload the perception results of CAVs to the edge to achieve shorter latency.
The above edge-assisted approaches generally neglect collaboration between CAVs, and the edge may become a transmission bottleneck when the number of CAVs engaging in collaborative perception is relatively large.
Subsequent studies further examined these issues.
Zaki et al.\cite{2024itvt-zaki-qualityawaretaskoffloading} proposed a quality-aware task offloading scheme that reduces perception redundancy by predicting vehicle motion to estimate its region of interest.
Although this approach demonstrates promising efficiency gains, its effectiveness depends on estimation accuracy, which may cause variability in dynamic traffic scenarios.
Lu et al.\cite{2025itvt-lu-jointoptimizationcompression} proposed a joint optimization framework for compression, transmission, and computation to minimize collaboration latency under perception and queue constraints.
While effective in addressing resource allocation, the framework is limited to pairwise vehicle interactions, which may hinder scalability in dense networks with multiple CAVs contending for the same provider.
Liang et al.\cite{2025ITMC-liang-FullPerceptionNetworklevelCollaborative} proposed scheduling semantic feature sharing in blind spots to enhance perception under communication constraints.
However, it is limited to local perception, potentially restricting its effectiveness in global awareness.

\section{The Proposed LGCP Framework}

\begin{figure*}[t!]
	\centering
	\includegraphics[width=0.9\linewidth]{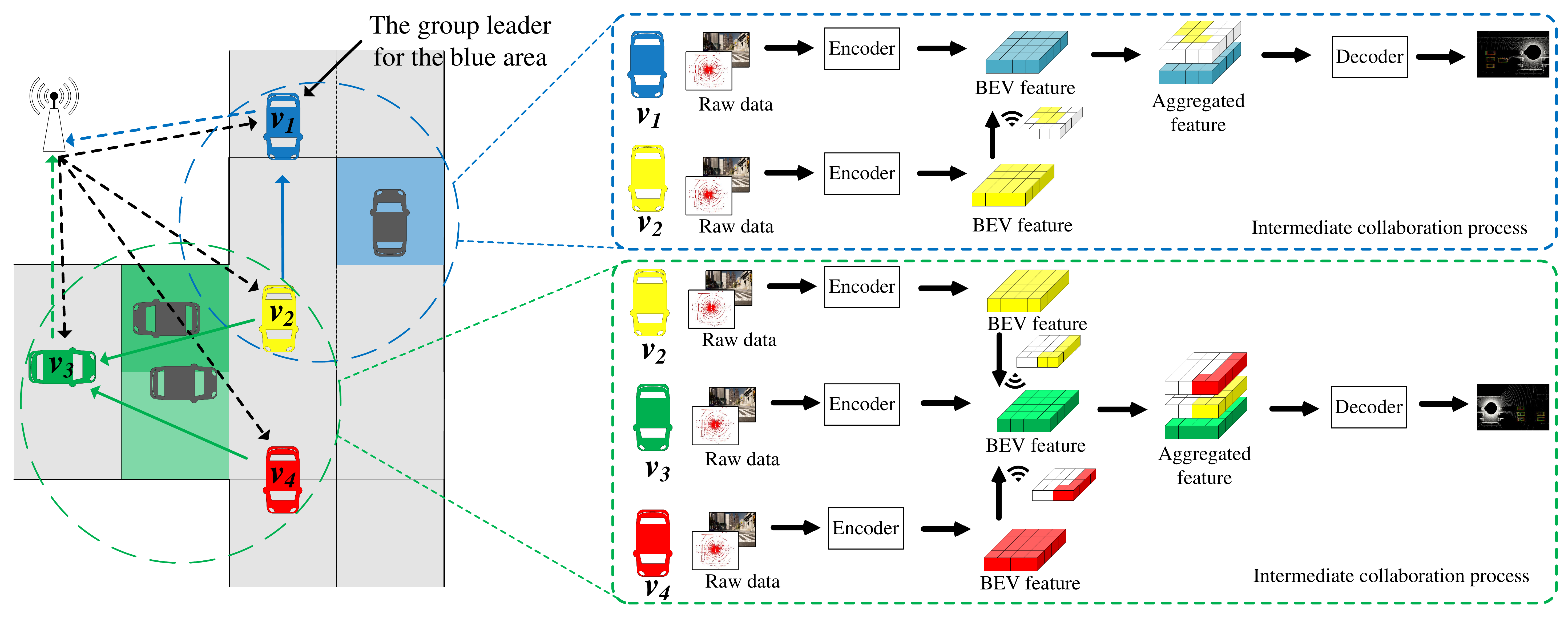}
        \caption{Object detection architecture of LGCP.  }
	\label{fig:architecture}
\end{figure*}

% We now propose the details of LGCP framework.
The LGCP framework adopts intermediate collaboration to balance perception performance and communication efficiency.
It employs a centralized scheduling strategy to achieve efficient task assignment and perception result aggregation. 
Compared to edge servers, RSUs are lighter and easier to deploy.
Thus, we focus on an RSU-assisted collaborative perception scenario, where the RSU acts as the central controller.
The LGCP framework mainly consists of the following steps: 
\begin{enumerate}
    \item Initiation. 
    The RSU partitions the RoI into a series of non-overlapping areas based on geographic boundaries of the RoI, and broadcasts an initiation message to start a global collaborative perception process.
    Upon reception of the initiation message, each CAV sends its basic information, such as location and direction to the RSU.
    \item Task assignment.
    The RSU assigns and broadcasts the perception task of each area to a CAV group.
    For each area, the RSU designates a group leader to collect and fuse perception data from its members.
    \item Data sharing and fusion. 
    CAVs partition their perception data into different areas according to the partitioning rule of the RSU, and then transmit the area-specific perception data to the corresponding group leader based on their tasks. 
    Group leaders then perform perception tasks for their corresponding areas, and upload the perception results to the RSU.
    \item Result aggregation and propagation. 
    The RSU constructs a global view based on the received perception results, and propagates the global view back to all participating CAVs.
\end{enumerate}
The above procedure operates in a continuous loop.
During each cycle, CAVs upload their basic information as part of the result aggregation phase, while the RSU simultaneously propagates the global view and the latest task assignment to all participating CAVs.

The LGCP framework incorporates various collaboration models for the purpose of object detection.
Fig. \ref{fig:architecture} shows the object detection architecture of the proposed LGCP framework.
The object detection task for the area marked in blue is assigned to CAVs $v_1$ and $v_2$, with $v_1$ designated as the group leader.
CAVs $v_1$ and $v_2$ employ an intermediate collaborative perception model, while $v_2$ extracts BEV features from its local raw sensor data using an encoder network and shares these features with $v_1$.
Subsequently, $v_1$ aggregates the received features and generates the perception result through a decoder network.
Similarly, CAVs $v_2$, $v_3$ and $v_4$ collaboratively perceive objects in areas marked in green, and $v_3$ is the corresponding group leader.
By decomposing the whole collaboration task into multiple area-based tasks and distributing them to multiple CAV groups, our design avoids the large-scale data aggregation problem and the centralized data processing problem.
% The inherent reduction in transmissions further mitigates the risk of link degradation.
% With the advent of 5G-V2X technology, CAVs and the edge server periodically broadcast their status information.
% As a result, only minimal additional data needs to be included in these routine broadcasts.
% Moreover, the high-frequency and low-latency nature of 5G-V2X communication enables accurate vehicle localization even in highly dynamic environments.
% Consequently, the edge server can monitor states and promptly exclude faulty CAVs, thereby ensuring the stability and reliability of collaboration. 

\section{Problem Formulation}
Perception accuracy and latency are two fundamental performance indices for assessing the performance of collaborative perception systems.
We first consider how to estimate perception accuracy.
Let $N$ denote the total number of areas within the RoI, $a_i$ denote the $i$th area, and $P_{acc}(a_i)$ denote the perception accuracy of area $a_i$. 
The final global perception accuracy can be represented as the average perception accuracy of all areas: $\sum_{i=1}^{N} P_{acc}(a_i) / {N}$.
For each area $a_i$, the value of $P_{acc}(a_i)$ depends on the CAV group assigned to $a_i$ and the quality of the perception data of each CAV in the group.
Since CAVs do not transmit their perception data to the RSU, it is hard for the RSU to accurately estimate the value of $P_{acc}(a_i)$ before an actual collaborative perception process starts. 
For such a reason, we introduce a metric called area confidence to approximately estimate the value of $P_{acc}(a_i)$. 
Each CAV first extracts intermediate features from its local raw data, and then evaluates the expected confidence level of its perception results for different areas based on these extracted features.
Let $f_{i,j}$ represent the feature of area $a_i$ extracted by CAV $v_j$, and the area confidence $F_i({\{v_j\}})$ can be calculated as:
\begin{equation} \label{eq:confidence_one_node}
    F_i({\{v_j\}}) = f_{gen}(f_{i,j}),
\end{equation}
where $f_{gen}$ is a decoding module\cite{2022ANIPS-hu-Where2commCommunicationEfficientCollaborative}.
Furthermore, let $\hat V_i$ represent the CAV group assigned to area $a_i$.
The collaborative area confidence $F_i(\hat V_i)$ is estimated as:
\begin{equation} \label{eq:block_confidence}
F_i(\hat V_i) = 1 - \prod_{v_k \in \hat V_i}(1-F_i({\{v_k\}})).
\end{equation}
Thus, the global perception accuracy is formulated as:
\begin{equation} \label{eq:global_confidence}
\frac {1} {N}\sum_{i=1}^{N}  {P_{acc}(a_i)}  \approx \frac {1} {N} \sum_{i=1}^{N} {F_i(\hat V_i)}.
\end{equation}

Next, we consider the calculation of the latency.
The overall latency mainly consists of two parts: transmission latency and computation latency.
Similarly to previous work \cite{2023IJSAC-luo-EdgeCooperNetworkAwareCooperative},  LGCP utilizes 5G-V2X sidelink for V2V communication, and 5G uplink and downlink for V2I communication.
The RSU assigns and manages wireless resources for V2X communications via centralized scheduling.

% The latency $t^{s_0}$ in initiation stage primarily stems from basic message exchange between the RSU and CAVs; 
In the first stage, the RSU broadcasts an initiation message to all CAVs through 5G downlink, and $|V|$ CAVs report their information to the RSU through 5G uplink to engage in collaboration. 
Since all CAVs share $Z$ subchannels, the expected transmission latency is approximately 
    $t_1 = \frac {D_{init}} {R_t} + \lceil \frac {|V|} {Z} \rceil * \frac {D_{info}} {R_t}$,
where $D_{init}$ is the length of the initiation message, $D_{info}$ is the length of the report message of a CAV, and $R_t$ is the data transmission rate.

In the second stage, the RSU broadcasts the task assignment results to all CAVs through 5G downlink, and the expected transmission latency is approximately 
    $t_2 = \frac {D_{ts}} {R_t}$,
where $D_{ts}$ is the length of the task assignment message.

In the third stage, all CAVs first transmit the area-specific perception data to the corresponding group leaders based on the transmission schedule scheme provided by the RSU.
Each group leader then fuses all received perception data and uploads the perception result to the RSU. 
The latency of area $a_i$ can be approximately calculated as $t^a(a_i) + t^f(a_i) + \frac{D_{rep} } {R_t}$, where $t^a(a_i)$ and $t^f(a_i)$ represent the data aggregation latency and the fusion latency of area $a_i$ respectively, and $D_{rep} $ represents the length of the report message.
Thus, the latency in this stage can be calculated as:
\begin{equation} \label{eq:t3}
    \begin{aligned}
    t_3 
    &= \max\limits_{1\leq i \leq N}(t^a(a_i) + t^f(a_i) +  \frac{D_{rep}} {R_t})\\
    &= \max\limits_{1\leq i \leq N}(t^a(a_i) + t^f(a_i)) +  \frac{D_{rep}} {R_t}\\
    &= |\mathcal{S}(\hat V)| +  \frac{D_{rep}} {R_t},
\end{aligned}
\end{equation}
where $\mathcal{S}(\hat V)$ is the adopted transmission schedule policy for $\hat V$, and $\hat {V} = \{\hat V_1,\hat V_2, ...,\hat V_{N}\}$ is the set of CAV groups for all areas.

In the fifth stage, the RSU broadcasts the global view to all CAVs, and the expected transmission latency is approximately $t_4 = \frac {D_{G}} {R_t}$, where $D_{G}$ is the length of the global view message.

As a result, the total latency can be approximately calculated as:
\begin{equation} \label{eq:total_latency}
    \begin{aligned}
    \sum_{k=1}^{4} t_i 
    &\approx \frac {D_{init} + \lceil \frac {|V|}{Z} \rceil * {D_{info}} + D_{ts} + D_{rep} + D_{G}} {R_t} + |\mathcal{S}(\hat V)|\\
    &= t_{\Delta} + |\mathcal{S}(\hat V)| ,
    \end{aligned}
\end{equation}
where $t_{\Delta} = \frac {D_{init} + \lceil \frac {|V|}{Z} \rceil * {D_{info}} + D_{ts} + D_{rep} + D_{G}} {R_t}$.

For each area $a_i$, increasing the size of $V_i$ generally improves perception accuracy, though it may also result in longer transmission and computation latency. 
To jointly balance the trade-off between perception accuracy and latency, we attempt to maximize the value of ${\frac {1}{N}\sum_{i=1}^{N} P_{acc}(a_i) } / {\sum_{i=1}^{4} t_i}$.
Substituting (\ref{eq:global_confidence}) and (\ref{eq:total_latency}) into ${\frac {1}{N}\sum_{i=1}^{N} P_{acc}(a_i) } / {\sum_{i=1}^{4} t_i}$, we obtain 
\begin{equation}
    \frac {\frac {1}{N}\sum_{i=1}^{N} P_{acc}(a_i) } {\sum_{i=1}^{4} t_i} \approx \frac {\frac {1}{N}\sum_{i=1}^{N} F_i(\hat V_i) } {t_{\Delta} + |\mathcal{S}(\hat V)|}.
\end{equation}
Thus, our optimization problem can be formalized as:
\begin{align}
    P_0: \quad &  {<\hat V^*, \mathcal{S}^*(\hat V^*)>}  =  \mathop{\arg \max} \limits_{<\hat {V}, \mathcal{S}>}\frac {\frac {1}{N}\sum_{i=1}^{N} F_i(\hat V_i) } {t_{\Delta} + |\mathcal{S}(\hat V)|}, \label{eq:problem0}\\
   \text{s.t.}  
    % & \quad\frac {1}{N}\sum_{i=1}^{N} f(V_i) \ge P_0, \tag{\ref{eq:problem0}{a}}\label{eq:problem0a}\\
    % & \quad \sum_{k=1}^{4} t_i \leq T, \tag{\ref{eq:problem0}{a}}\label{eq:problem0a} \\
    & \quad t_{\Delta} + |\mathcal{S}(\hat V) |\leq T, \tag{\ref{eq:problem0}{a}}\label{eq:problem0a} \\
    & \quad \hat V_i \subset V, 1\leq i \leq N, \tag{\ref{eq:problem0}{b}}\label{eq:problem0b}
    \end{align}
where $T$ is the predefined maximum latency, and (\ref{eq:problem0a}) is designed to support dynamic scenarios by enforcing the real-time constraint, thereby ensuring responsiveness to vehicle mobility and system stability.

\section{Solution}
According to the definition of the optimization problem $P_0$, the solution is subject to two sub-optimization problems: selection of group sets $\hat {V}$ and the transmission schedule policy $\mathcal{S}(\hat {V})$.
In this section, we first consider the selection of group sets $\hat {V}$, and then consider the transmission schedule policy $\mathcal{S}(\hat {V})$.

\subsection{Selection of Group Sets $\hat {V}$}

Given a CAV set $V$, we select a group $\hat V_i \subset V$ for each area $a_i$ such that the confidence $F_i(\hat V_i)$ is as high as possible. 
In general, the confidence $F_i(\hat V_i)$ can be enhanced by incorporating more CAVs into the group $\hat V_i$.
However, the confidence gain from additional CAVs decreases as the group grows, as we can observe from Eq. (\ref{eq:block_confidence}), while expanding the group also imposes an inherent burden of both transmission and fusion.
For such a reason, we introduce a confidence incremental threshold $\Delta_g$ to limit the size of the group, and add a new CAV $v_j$ into the group $\hat V_i$ when the corresponding confidence gain is larger than or equal to the predefined confidence incremental threshold $\Delta_g$:
\begin{equation} \label{eq:Delta_g}
    F_i(\hat V_i \cup \{v_j\}) - F_i(\hat V_i) \ge \Delta_g.
\end{equation}

Upon obtaining the CAV group of each area, the next step is to designate a leader CAV for each group.
Let $y_{i,j}$ represent whether $v_j$ is the leader CAV of group $\hat V_i$.
Since each group has only one leader CAV, we obtain 
\begin{equation} \label{eq:block_yij}
    \forall \hat V_i \in \hat V, \sum_{v_j \in \hat V_i}y_{i,j} = 1.
\end{equation}
Since it is relatively straightforward to homogenize the heterogeneous perception features from CAVs~\cite{2025APA-gao-STAMP}, we assume that all area-specific features have the same size $B$. 
% In our formulation, a group is determined by both its member CAVs and its designated leader, and a CAV may belong to multiple groups and serve as the leader of several groups.
The fusion load $L_j$ of $v_j$ can be calculated as:
\begin{equation} \label{eq:node_load_process}
    L_j = \sum_{i=1}^{N}y_{i,j} |\hat V_i|B.
\end{equation}

In our framework, groups covering the same CAVs but related to different areas can have distinct leaders, and a CAV may belong to multiple groups and serve as the leader of several groups.
Since the data fusion process of each area operates independently of the others, evenly distributing the data fusion tasks to different CAVs can efficiently exploit the computation power of CAVs, thus decreasing the fusion latency.
Hence, we attempt to minimize the max task load of all CAVs: $\min \max_{v_j \in V} L_j$.

To solve this min-max problem, we employ a greedy strategy to derive an approximate solution and develop a selection algorithm.
The pseudocode of the proposed selection algorithm is presented in Algorithm \ref{alg:sel}. 
The codes from lines \ref{line:arg_sel_0} to \ref{line:arg_sel_1} first sort $V$ based on the corresponding area confidence and then sequentially select groups for different areas according to Eq. (\ref{eq:Delta_g}).
The codes from lines \ref{line:arg_sel_2} to \ref{line:arg_sel_3} assign a leader CAV to each group in sequence, such that $\hat{V}_i$ now includes information about its designated leader CAV.

\begin{algorithm}
\caption{Selection Algorithm}\label{alg:sel}
\textbf{Input:} CAV set $V$, confidence incremental threshold $\Delta_g$. 

\textbf{Output:} group set $\hat V$.

\begin{algorithmic}[1]
\State  $\hat{V} \leftarrow \emptyset$; 
\For{each perception area $a_i$} \label{line:arg_sel_0}
    \State Construct group $\hat{V}_i \subseteq V$ greedily based on Eq.(\ref{eq:Delta_g}).
    \State $\hat{V} \leftarrow \hat{V} \cup \{\hat V_i\}$;
\EndFor \label{line:arg_sel_1}

% \Statex \textbf{Assign aggregators:} 
\State Sort $\hat V_i \in \hat{V}$ in descending order of $|\hat V_i|$;\label{line:arg_sel_2}
\For{each group $\hat{V}_i \in \hat{V}$}  
    \State Select $v_k \in \hat{V}_i$ with minimal load $L_k$.
    \State Assign $v_k$ as the leader CAV for $a_i$, and update $L_k$.
\EndFor \label{line:arg_sel_3}
% \State Return $\hat{V}=\{\hat{V}_i\}$
\end{algorithmic}
\end{algorithm}

\textbf{Complexity analysis:} 
The time complexity of constructing groups is dominated by sorting and greedy selection for each area, resulting in $O(N |V| \log |V|)$. 
The leader CAV assignment involves sorting the $N$ groups by size, costing $O(N \log N)$, and selecting the minimal-load CAV for each group, taking $O(N |V|)$. 
Therefore, the overall complexity is $O(N |V| \log |V| + N \log N)$.

\subsection{Transmission Schedule Policy $\mathcal{S}(\hat {V})$}
We assume a 5G-V2X channel is divided into $Z$ logical subchannels, collectively denoted as a set $\hat{Z}$.
All subchannels are shared among all V2X links, and each link is restricted to a single subchannel allocation.
We assume each CAV is equipped with only one transmitter, and all CAVs operate in half-duplex mode and cannot perform transmission and reception operations simultaneously.
In wireless communications for 5G V2X links, two distinct types of interference constrain the operation of the system: self-interference and co-channel interference.
Self-interference represents that transmission nodes cannot simultaneously function as both source and destination, while co-channel interference represents that if a receiver associated with one transmitter falls within the interference range of another transmitter, the two transmitters cannot share the same channel for transmission.

% For each area $a_i$, all CAVs in $\hat V_i$ except the designated aggregator send their perception data to the aggregator.

For each area $a_i$, all CAVs in $\hat V_i$ other than the designated leader send their perception data to the leader.
We assume each CAV uses a data packet $p$ to encapsulate the perception data of a specific area.
The parameters of packet $p$ are represented as a 5-tuple: $\langle v_s,v_r,a,z,t \rangle$, where $p.v_s$ and $p.v_r$ represent the source and destination addresses of $p$, respectively, $p.a$ represents the corresponding area of $p$, $p.z$ represents the subchannel to be used for transmission of $p$, and $p.t$ represents the transmission moment.
Let $P$ denote the set of packets awaiting transmission, which can be initialized based on $\hat V$, and $l_{s\rightarrow r}$ represent a directed link between $v_s$ and $v_r$.
We adopt a binary variable $I_E(p)$ to indicate whether link $l_{s \rightarrow r}$ of $p$ conflicts with the transmission of any other scheduled packets in scheduled set $E$ according to the self-interference and co-channel interference rules.

The transmission order of packets in $P$ directly affects the value of $\mathcal{S}(\hat V)$.
Since the fusion processes of different leader CAVs are independent of each other, a leader CAV can fuse received packets during transmission once all packets are fully received. 
Therefore, parallelizing the data aggregation process and data fusion process can help to reduce the value of $\mathcal{S}(\hat V)$.
Thus, we utilize a priority metric $\omega(v_s,v_r)$ associated with packet $p$ that evaluates the joint impact of sender and receiver loads on network congestion and fusion latency.
Let $L_s(v_s)$ denote the sender load (number of packets transmitted from $v_s$) of $v_s$, and $L_r(v_r)$ denote the receiver load (number of packets  received at $v_r$).
The priority of packet from $v_s$ to $v_r$ formulated as:
\begin{equation} \label{eq:pri}
    \omega(v_s,v_r) = L_s(v_s) + L_r(v_r).
\end{equation}
Depending on this metric, we establish a scheduling order for packets.
% Thus, the transmission and fusion process can be regarded as a two-stage flow shop problem.

% Based on assigned clusters, the amount of data of $a_i$ can be calculated as $|\hat V_i|B_f$, and the corresponding transmission latency and fusion latency are $\frac{|\hat V_i|B_f}{R_t}$ and $\frac{|\hat V_i|B_f}{R_f}$, respectively, where $R_f$ is the fusion rate calculated as the required GPU floating-point operations per bit divided by the computational power (in the unit of Floating Point Operations per Second, FLOPS)\cite{2025itvt-lu-jointoptimizationcompression}.
% Johnson's Rule provides an efficient ordering method for the two-stage flow shop problem\cite{1954nrlq-johnson-optimaltwoandthreestage}, which accounts for both transmission latency and fusion latency of all areas.
% The rule works by iteratively selecting the area with the shortest processing latency, whether it is transmission latency or fusion latency, from the remaining unordered blocks.
% The selected area is then placed either at the beginning of the sequence if the shortest latency occurs in transmission or at the end if the shortest latency occurs in fusion.
% This process continues until all areas have been ordered.
% Following Johnson's Rule, we establish a processing order for packets.

Based on the above analysis, we develop a transmission scheduling algorithm.
The pseudocode of the proposed scheduling algorithm is presented in Algorithm \ref{alg:trans}.
The time slot $\tau$ is set to the duration required to transmit one packet.
The codes from lines \ref{line:arg_trans_2} to \ref{line:arg_trans_3} select the scheduled packet based on the priority and interference.
In Line \ref{line:arg_trans_4}, we calculate $\mathcal{S}(\hat V) $ based on the latest transmission time and the maximum remaining fusion time.

\begin{algorithm}[ht]
\caption{Transmission Scheduling}\label{alg:trans}
\textbf{Input:} subchannel set $\hat{Z}$, time slot $\tau$, group set $\hat{V}$.  

\textbf{Output:} joint latency $|\mathcal{S}(\hat{V})|$.

\begin{algorithmic}[1]
\State Convert the selection results of group set $\hat{V}$ into packet set $P$ for transmission;
\State Establish a scheduling order for packets based on Eq.(\ref{eq:pri}).
% \Statex \textbf{Determines processing order:}  
% \State Partition $P$ into two logically distinct parts $P_1$ and $P_2$, where $P_1$ is placed before $P_2$.  \label{line:arg_trans_0}
% \State Sort $P_1$ by increasing $\frac{|\hat{V}_{p.b}|}{R_t}$; sort $P_2$ by decreasing $\frac{|\hat{V}_{p.b}|}{R_f}$.\label{line:arg_trans_1}
 
\While{$ P \neq \emptyset$} \label{line:arg_trans_2}
    \State \(E \leftarrow \emptyset\);
    \For{each subchannel $z \in \hat{Z}$}
        \State Select a packet $p \in P$ that ensures $I_E(p)=0$;
        \If{$p$ exists}
            \State $p.z \leftarrow z$, $p.t \leftarrow t$;
            \State $P \leftarrow P \setminus \{p\}, E \leftarrow E \cup \{p\}$;
        \EndIf
    \EndFor
    \State Update load and remaining fusion time of each CAV upon full reception of all packets for an area.
    \State $ t  \leftarrow t  + \tau $; 
\EndWhile \label{line:arg_trans_3}

\State Return the joint latency $|\mathcal{S}(\hat{V})|$ based on the latest transmission time and the maximum remaining fusion time.\label{line:arg_trans_4}
\end{algorithmic}
\end{algorithm}

\textbf{Complexity analysis:}
The complexity of the conversion operation is only $O(|P|)$.
The complexity of determining processing order is $O(|P| \log |P|)$ due to the sorting operations.
The scheduling process requires $O(|P|)$ time per packet in the worst case due to the interference constraints.
This results in a total complexity of $O(|P|^2)$.

\subsection{Overall Algorithm}

Building on the previously described algorithms, Algorithm \ref{alg3} presents the pseudocode of our proposed solution.
The algorithm operates through a two-stage process: first, it employs Algorithm \ref{alg:sel} to obtain the group set $\hat V$, and then employs Algorithm \ref{alg:trans} to schedule the transmission and fusion process.

\textbf{Complexity analysis:} The overall computational complexity comprises two interconnected components.
Algorithm \ref{alg:trans} processes the output from Algorithm \ref{alg:sel} as its input.
Considering that the input size $|P|$ of Algorithm \ref{alg:trans} scales with the number of areas $N$, the computational complexity of Algorithm \ref{alg:trans} is $O(N^2)$, and the overall computational complexity is $O(N|V|\log|V|+N^2)$.

\begin{algorithm}[H]
\caption{Overall Algorithm}\label{alg3}
\textbf{Input:} CAV set ${V}$, confidence incremental threshold $\Delta_g$, subchannel set $\hat{Z}$, time slot $\tau$.

\textbf{Output:} joint latency $\mathcal{S}^*(\hat{V})$.

\begin{algorithmic}[1]
\State  Employ Algorithm \ref{alg:sel} to obtain the group set $\hat V$;
\State  Employ Algorithm \ref{alg:trans} to schedule the transmission and fusion process.
\end{algorithmic}
\end{algorithm}

\section{EXPERIMENT}
In this section, we evaluate the detection performance of various collaboration models integrated with our LGCP framework, the amount of data transmission and the end-to-end latency for various mechanisms.
\subsection{Dataset and Simulation Platform}
\textbf{OPV2V}\cite{20222ICRAI-xu-OPV2VOpenBenchmark} is a simulated V2V collaborative perception dataset with 73 driving scenarios, including 11464 frames of 3D point clouds. 
Each scenario contains some traditional vehicles and several CAVs (2 to 7). 
Each CAV is equipped with LiDAR, while traditional vehicles are not.
% The simulated LiDAR is streamed at 20 Hz and recorded at 10 Hz. 
% The training/validation/testing splits include 6,764, 1,981, and 2,719 frames, respectively.

\textbf{V2XSet}\cite{2022ECCV-xu-V2XViTVehicletoEverythingCooperative} is a simulated V2X dataset supporting vehicle-to-everything perception. 
It includes 5 representative scenarios covering 5 different roadway types across 8 towns in CARLA\cite{2017CRL-dosovitskiy-CARLAOpenUrban}, with a total of 11447 frames. 
Each scenario contains some traditional vehicles, and at least 2 and at most 7 intelligent agents (CAV or infrastructure). 
Each agent is equipped with LiDAR, while traditional vehicles are not.
% The simulated LiDAR is recorded at 10 Hz and saves the corresponding positional data and timestamps. 
% The training/validation/testing sets are 6,694, 1,920, and 2,833 frames, respectively.

To validate our approach, we employ the OpenCOOD\cite{20222ICRAI-xu-OPV2VOpenBenchmark} simulation platform. 
% We evaluate the performance of the object detection task on the testing sets, and calculate the average overhead in scenarios with a total of 7 CAVs, for varying numbers of CAVs participating in the collaboration.
The infrastructure in V2XSet is treated as a fixed CAV for evaluation.
In addition to the evaluation on datasets, we employ a co-simulation platform that integrates the autonomous driving simulator CARLA, collaborative driving automation framework OpenCDA\cite{xu2021opencda} and the network simulator NS3\cite{carneiro2010ns} to further conduct experiments in multi-CAV deployment scenarios.

\subsection{Baseline and Metric}
To comprehensively assess the impact of the proposed LGCP framework on collaboration models, we implement and evaluate three collaboration models: CoBEVT\cite{2022APA-xu-CoBEVTCooperativeBirds}, Where2comm\cite{2022ANIPS-hu-Where2commCommunicationEfficientCollaborative}, and CoAlign\cite{20232IICRAI-lu-RobustCollaborative3D}.
We also examine the model performance under a single-vehicle setting, named no collaboration, which excludes V2V communication. 
To quantitatively assess the communication efficiency of the LGCP framework, we evaluate the amount of data transmission and end-to-end latency, compared with the vehicle-based paradigm and the edge-assisted paradigm. 
To further assess the effectiveness, we compare it with the proactive
conflict-free scheduling (PCS) of Fullperception\cite{2025ITMC-liang-FullPerceptionNetworklevelCollaborative} and a random scheduling baseline in multi-CAV deployment scenarios.
To evaluate the performance of the LGCP framework, we consider the following performance metrics:
\begin{itemize}
    \item[$\bullet$] \textit{Average precision}. It is calculated by measuring the precision-recall curve at different Intersection-over-Union (IoU) thresholds between the predicted bounding boxes and ground truth annotations. Specifically, we compare three IoU thresholds of 0.3, 0.5, 0.7. This metric reflects the object detection performance.
    \item[$\bullet$] \textit{Amount of data transmission}. It is defined as the total amount of data transmitted by all CAVs to complete a single collaborative perception. This metric reflects the overhead associated with the collaborative perception task.
    \item[$\bullet$] \textit{End-to-end latency}. Low end-to-end latency is crucial for real-time collaborative perception. It is measured as the time interval from the initiation to the completion of a collaborative perception task. 
\end{itemize}

% \begin{figure}[ht]
%     \centering
%     \includegraphics[width=0.45\linewidth]{picture/chunk1.pdf}
%     \caption{Supplementary transmission based on chunks.}
%     \label{fig:chunk}
% \end{figure}

\subsection{Experimental Settings}
Following \cite{20222ICRAI-xu-OPV2VOpenBenchmark,2022ECCV-xu-V2XViTVehicletoEverythingCooperative}, we set the detection range as $280m \times 80m$, where all the CAVs are included in this spatial range in the experiment.
For a fair comparison, all models employ PointPillar\cite{2019PICCVPR-lang-PointPillarsFastEncoders} as the backbone with a 10ms delay\cite{20222ICRAI-xu-OPV2VOpenBenchmark} and maintain consistent parameters for each type of model.
The integrated models maintain the same feature dimensions as the original models within the LGCP framework.
The computational capacities for data fusion are configured as 0.1 tera floating-point operations per second (TFLOPS) for CAVs and 2 TFLOPS for the edge server, respectively.
The RoI is divided into grid areas of size $10\text{m} \times 6\text{m}$, each of which is approximately twice the length of a typical car and twice the width of a standard lane.
Areas are adaptively represented by grids that are currently occupied by vehicles.
A 5G-V2X channel is divided into five 8MHz subchannels.
% Data transmission in LGCP is disabled when the Signal-to-Interference-and-Noise Ratio (SINR) falls below a predefined threshold $\phi$, and resumes at a fixed transmission rate of 27 Mbps once the SINR reaches or exceeds this threshold.
% The data size of $D_{init}, D_{info}, D_{ts}, D_{rep}, D_G$ are set to 16kb, 4kb, 8kb, 4kb, and 22kb, respectively, with a total latency $t_\Delta$ of approximately 2ms.
% Each shared feature is compressed to 2.16 Mb for transmission, yielding a transmission latency of approximately $(\frac {10\times6} {280\times80}\times2.16\times10^3)/27 \approx 0.22ms$ per packet.
Each complete shared feature is compressed to 2.16Mb.
The corresponding costs in million floating point operations (MFLOPs) for the three models are 2228, 1400, and 2684, respectively.
In the evaluation based on datasets, data transmission is disabled when the achievable transmission rate falls below 27 Mbps. 
Once the rate exceeds this threshold, transmission resumes at a fixed rate of 27 Mbps.
In the vehicle-based paradigm, all CAVs concurrently require perception data from other CAVs for collaborative perception.
Collaboration is considered to be complete only after all CAVs finish computation.
% In the edge-assisted mechanism, CAVs can upload perception data to the edge server simultaneously.
In our setting, CAVs under both the vehicle-based paradigm and the edge-assisted paradigm transmit all the perception data for collaborative perception.
For the other transmission parameters, we follow the simulation of EdgeCooper\cite{2023IJSAC-luo-EdgeCooperNetworkAwareCooperative}.
The major parameters are listed in Table \ref{tab:table0}.

 \begin{table}[!t]
	\caption{SUMMARY OF TRANSMISSION PARAMETERS.\label{tab:table0}}
	\centering
        \setlength{\tabcolsep}{3mm}{
	\begin{tabular}{c|c}
            \hline
            Parameter & Value \\
		\hline \hline
            Frequency band & 5.9GHz \\
		  Bandwidth & 40MHz \\
            Numer of subchannels $Z$ & 5 \\
            Transmission power & 23dbm \\
            % Transmission rate  upper bound & 27Mbps \\
            CAV computational capability & 0.1TFLOPS\\
            % SINR threshold $\phi$ & 7.4dB\\
            Time slot $\tau$ & 0.25ms \\
            % $t_\Delta$ &2ms\\
            Maximum latency $T$ & 100ms \\
            % Propagation model & Omnidirectional propagation model\\
		Path loss model  & 128.1+36.6$\log_{10}(d)$ \\
            Shadowing distribution &Log-normal \\
            Shadowing standard deviation & 8dB \\
		Noise power & -114dbm \\
		\hline
	\end{tabular}
        }
\end{table}

\subsection{Quantitative Evaluation}

% \begin{figure*}[htbp]    
%   \centering           
%   \subfloat[OPV2V.]   
%   {    \label{fig:opv2v_g}\includegraphics[width=0.45\textwidth]{picture/g_op.pdf}
%   }
%   \subfloat[V2Xset.]
%   { \label{fig:v2xset_g}\includegraphics[width=0.45\textwidth]{picture/g_set.pdf}
%   }
%   \caption{AP@0.7 performance among three collaboration models integrated with LGCP framework and the amount of data transmission for varying $\Delta_g$.}   
%   \label{fig:g}           
% \end{figure*}

\begin{figure*}[htbp]
  \centering
  \begin{minipage}[t]{0.45\textwidth}
    \centering
    \includegraphics[width=\linewidth]{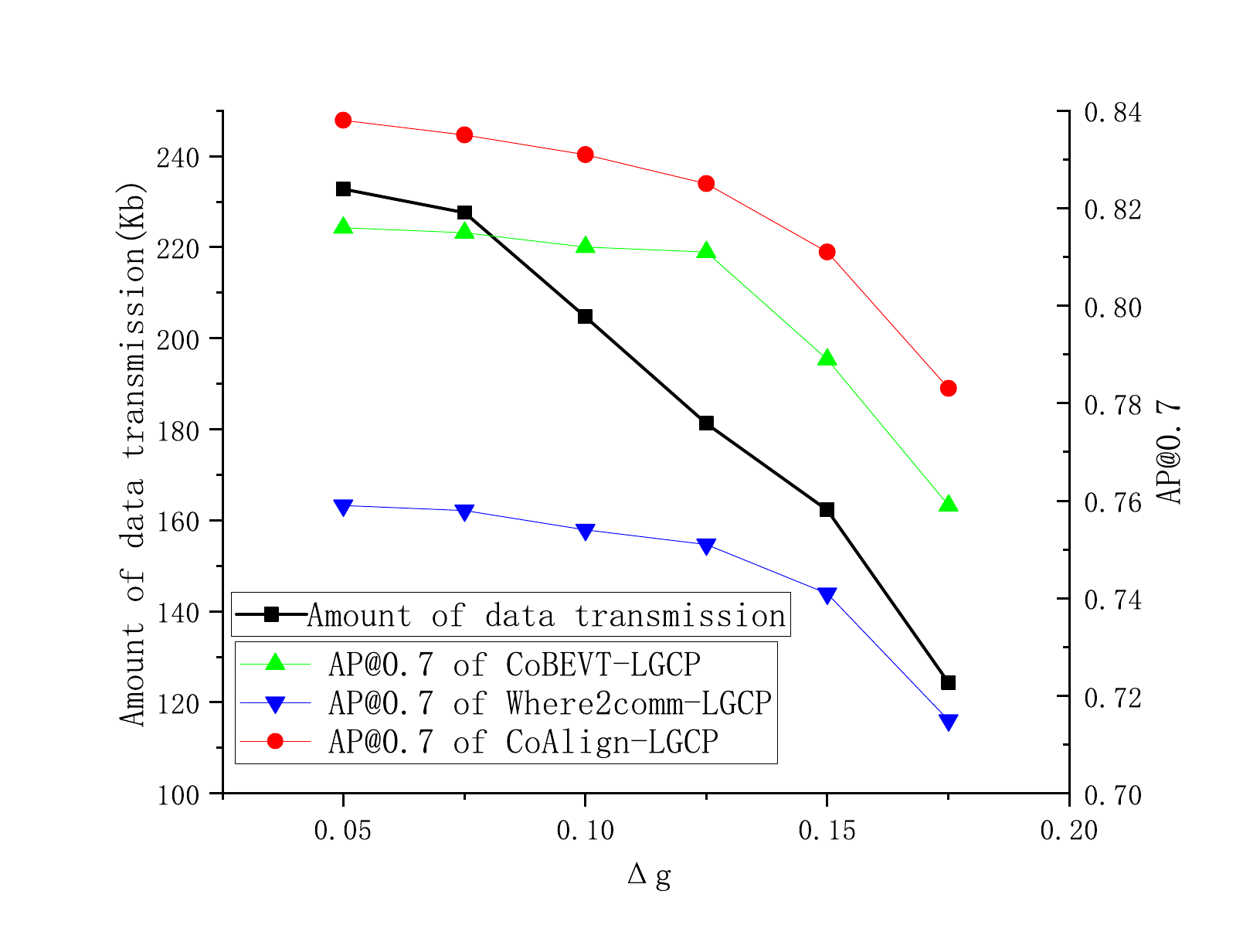}
    \vskip -3ex
    {\small (a) OPV2V.}
    \label{fig:opv2v_g}
  \end{minipage}
  \hfill
  \begin{minipage}[t]{0.45\textwidth}
    \centering
    \includegraphics[width=\linewidth]{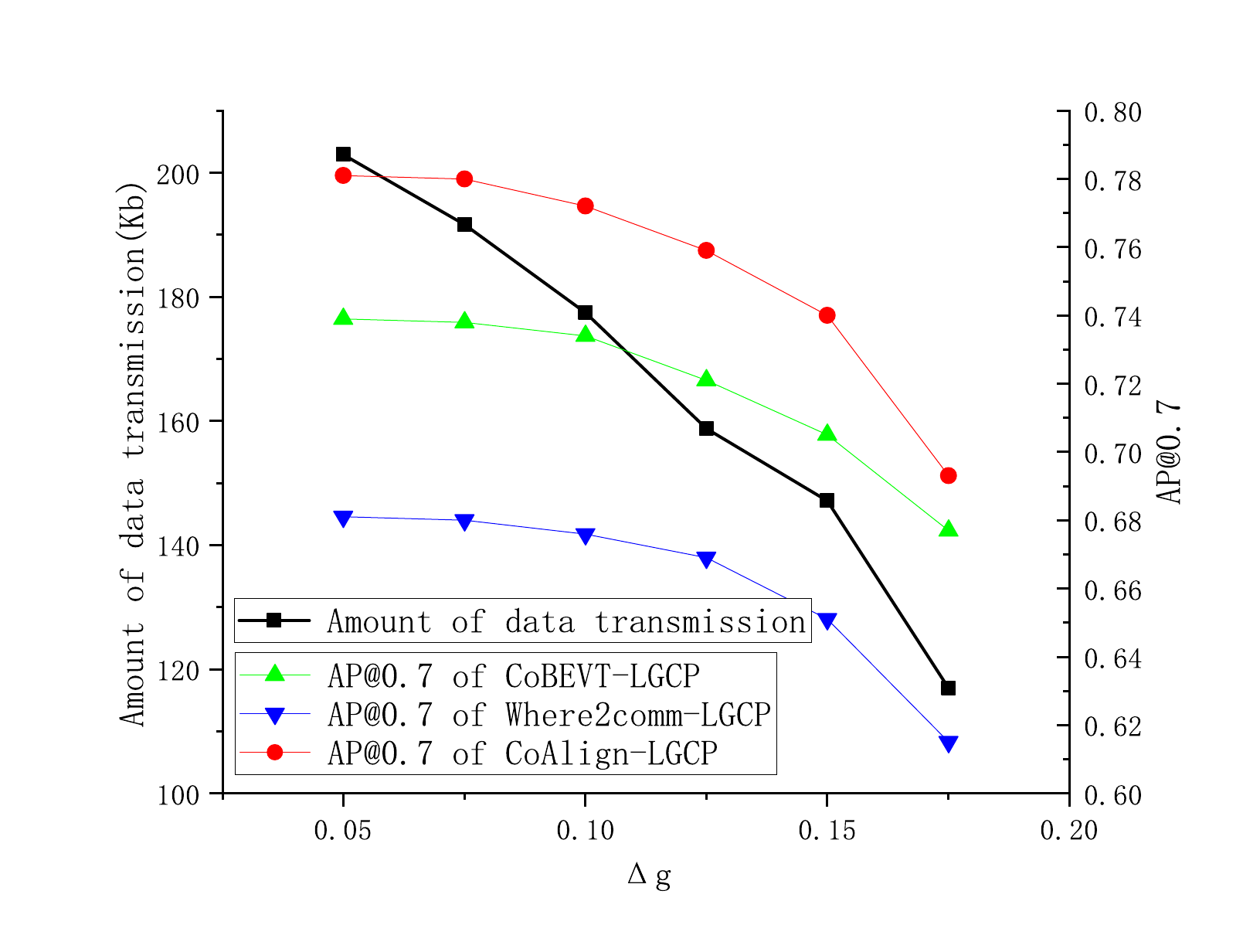}
    \vskip -3ex
    {\small (b) V2Xset.}
    \label{fig:v2xset_g}
  \end{minipage}
  \vskip -1.5ex
  \caption{AP@0.7 performance among three collaboration models integrated with LGCP framework and the amount of data transmission for varying $\Delta_g$.}
  \label{fig:g}
\end{figure*}
The group for each area may vary under different conditions of $\Delta_g$, which affects perception accuracy. 
Fig. \ref{fig:g} shows the average precision at IOU=0.7 (AP@0.7) performance among three collaboration models integrated with LGCP framework and the amount of data transmission for varying $\Delta_g$.
% demonstrates the impact varying $\Delta_g$ for selection of collaboration set on LGCP framework in two datasets.
% The left y-axis represents the total transmission volume measured by blocks for each collaboration on average and the right y-axis represents the corresponding AP@0.7.
% In both datasets, when $\Delta_g$ increases from $0.05$ to $0.10$, the perception accuracy is maintained at a comparable level. 
% However, when $\Delta_g$ increase from $0.10$ to $0.15$, the perception accuracy drop significantly.
As the $\Delta_g$ increases, the RSU tends to select fewer CAVs for each group, resulting in a decrease in the perception accuracy.
% When $\Delta_g$ rises above a specific value, the collaboration performance begins to decline significantly.
% Based on these findings, we set $\Delta_g$ to 0.1 for subsequent evaluations.
% At this time, the average transmission volume is approximately 66Kb in OPV2V and 50kb in V2XSet.

 \begin{table*}[!t]
	\caption{3D detection performance comparison. We show Average Precision (AP) at IoU=0.3, 0.5, 0.7 in no collaboration, collaboration models, and collaboration models integrated with LGCP framework.\label{tab:table1}}
	\centering
        \setlength{\tabcolsep}{3mm}{
	\begin{tabular}{c|ccc|ccc}
		\hline
            & \multicolumn{3}{c|}{\textbf{OPV2V}}  &\multicolumn{3}{c}{\textbf{V2XSet}}\\
            \cline{2-7}
		  & AP@0.3 & AP@0.5 & AP@0.7 & AP@0.3 & AP@0.5 & AP@0.7\\
		\hline
		No collaboration & 0.769 & 0.738 & 0.563 & 0.650 & 0.593 & 0.391\\
		\hline
            CoBEVT & 0.881 $\pm$ 0.03  & 0.879 $\pm$ 0.04 & 0.826 $\pm$ 0.03               & 0.833 $\pm$ 0.05 & 0.822 $\pm$ 0.04 & 0.738 $\pm$ 0.06\\
            % CoBEVT-mean & 0.887 &0.884 & 0.827\\ 
            CoBEVT-LGCP($\Delta_g=0.125$) & 0.878 & 0.875 & 0.811               & 0.822 & 0.808 & 0.721\\
            CoBEVT-LGCP($\Delta_g=0.1$) & 0.878 & 0.875 & 0.812               & 0.830 & 0.817 & 0.734\\
            CoBEVT-LGCP($\Delta_g=0.075$) & 0.878 & 0.875 & 0.815               & 0.829 & 0.817 & 0.738\\
            CoBEVT-LGCP($\Delta_g=0.05$) & 0.879 & 0.876 & 0.816               & 0.828 & 0.816 & 0.739\\
  	\hline
            Where2comm & 0.884 $\pm$ 0.07 & 0.868 $\pm$ 0.10 & 0.762 $\pm$ 0.09           & 0.816 $\pm$ 0.26 & 0.799 $\pm$ 0.28 & 0.651 $\pm$ 0.16 \\
            % Where2comm-mean &0.885 & 0.872 & 0.770 &0.822 & 0.805& 0.650\\
            Where2comm-LGCP($\Delta_g=0.125$) & 0.902 & 0.890 & 0.751           & 0.871 & 0.850 & 0.669\\
            Where2comm-LGCP($\Delta_g=0.1$) & 0.901 & 0.890 & 0.754           & 0.871 & 0.851 & 0.676\\
		Where2comm-LGCP($\Delta_g=0.075$) & 0.903 & 0.892 & 0.758           & 0.870 & 0.851 & 0.680\\
            Where2comm-LGCP($\Delta_g=0.05$) & 0.905 & 0.893 & 0.759           & 0.869 & 0.851 & 0.681\\
		\hline
            CoAlign & 0.899 $\pm$ 0.08 & 0.891 $\pm$ 0.10 & 0.818 $\pm$ 0.12              & 0.847 $\pm$ 0.23 & 0.832 $\pm$ 0.18  & 0.738 $\pm$ 0.31\\
            % CoAlign-mean & 0.900& 0.890 &0.820&0.847&0.831&0.718\\
            CoAlign-LGCP($\Delta_g=0.125$) & 0.924 & 0.916 & 0.825              & 0.893 & 0.879 & 0.759\\
            CoAlign-LGCP($\Delta_g=0.1$) & 0.923 & 0.916 & 0.831              & 0.897 & 0.883 & 0.772\\
		CoAlign-LGCP($\Delta_g=0.075$) & 0.924 & 0.918 & 0.835              & 0.900 & 0.888 & 0.780\\
            CoAlign-LGCP($\Delta_g=0.05$) & 0.925 & 0.919 & 0.838              & 0.899 & 0.887 & 0.781\\
		\hline
	\end{tabular}
        }
\end{table*}

To further evaluate the perception performance, we conduct a multi-threshold assessment by calculating the detection accuracy for varying IoU thresholds.
For each collaboration model, we evaluate the results with different CAVs as the ego CAV and represent the range in the form of midpoint$\pm$deviation.
Table \ref{tab:table1} summarizes the perception performance of no collaboration, collaboration models and collaboration models integrated with the LGCP framework. 
Experimental results show that the detection results relying solely on a single CAV are suboptimal.
In contrast, CAVs benefit significantly from complementary perception data through collaborative perception.
Compared with Where2comm and CoAlign, the deviation of CoBEVT is smaller when selecting different CAVs as the ego CAV.
It is because CoBEVT adopts a fused axial attention module to capture sparse local and global spatial interaction across views and CAVs, which makes CoBEVT exhibit consistent performance across different viewpoints.
% Compared to Where2comm and CoAlign, CoBEVT exhibits smaller deviation across ego CAVs, due to its fused axial attention module that captures sparse local and global spatial interactions across views and vehicles, leading to more consistent performance.
In most cases on both datasets, the LGCP framework improves the performance of Where2comm and CoAlign, especially in AP@0.3 and AP@0.5.
This performance enhancement can be attributed to inherent differences in viewpoints among CAVs at different positions.
Although CAVs receive complementary information from others, the acquired data may not be fully leveraged by the collaboration models.
This is why integrating the LGCP framework leads to performance improvements in Where2comm and CoAlign.
Meanwhile, although CoBEVT considers the relationships between different viewpoints during fusion, the performance of CoBEVT integrated with the LGCP framework still remains close to the original model, even though each CAV transmits only partial data instead of complete feature maps.
As $\Delta_g$ decreases, the impact on AP@0.3 and AP@0.5 is minimal, especially when $\Delta_g$ is below 0.1.
In contrast, the performance for AP@0.7 gradually improves, as smaller $\Delta_g$ values allow more perception data to be transmitted for collaboration.
However, when $\Delta_g$ is below 0.075, the rate of improvement becomes negligible.
Therefore, we set $\Delta_g$ to 0.075 for subsequent evaluations.

\begin{figure}
    \centering
    \includegraphics[width=0.9\linewidth]{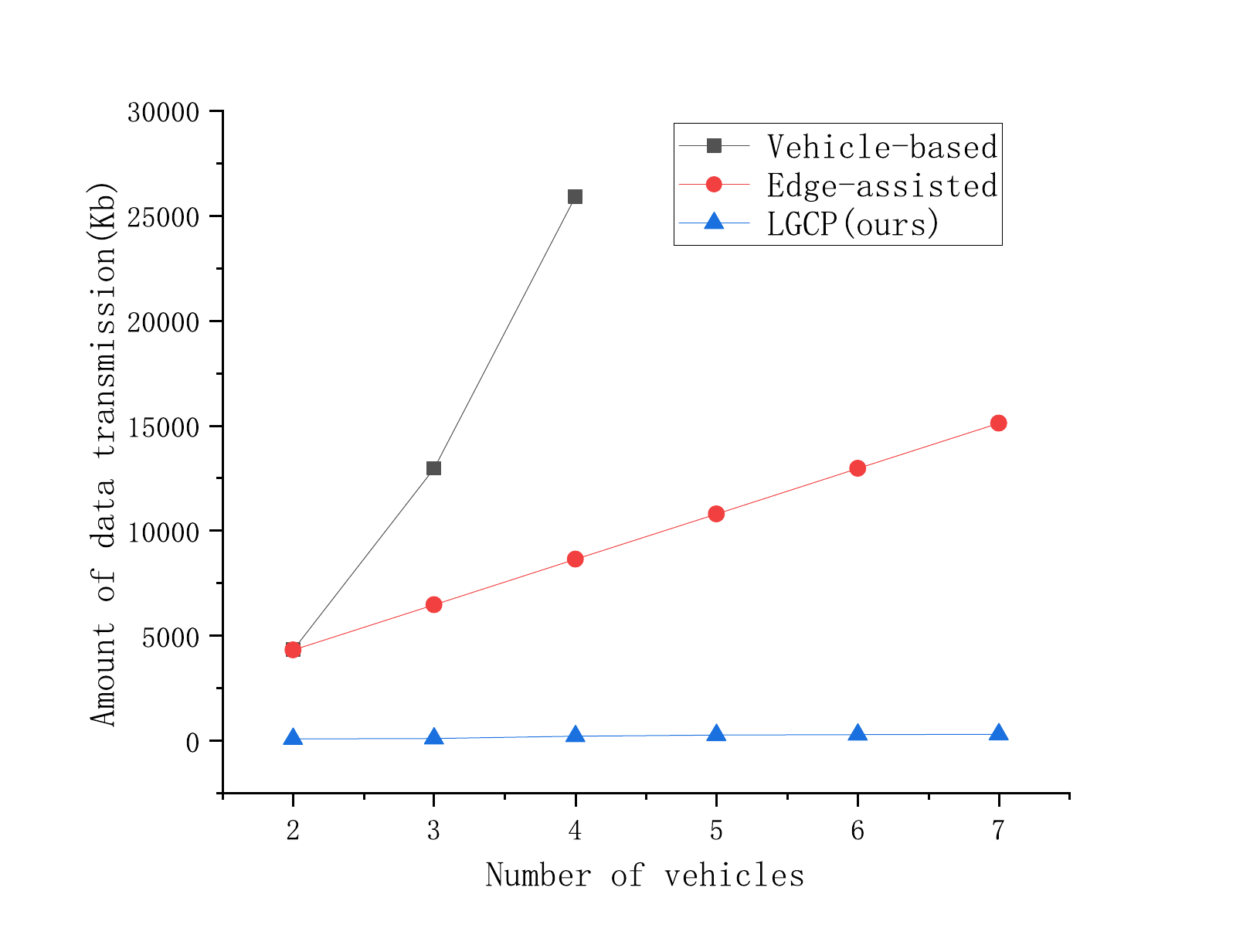}
    \caption{Amount of data transmission for varying number of CAVs under the OPV2V dataset.}
    \label{fig:volume}
\end{figure}

Fig. \ref{fig:volume} compares the amount of data transmission required for all CAVs to complete collaboration for three different schemes under the OPV2V dataset. 
In the vehicle-based paradigm, each CAV transmits perception data to all other CAVs to complete collaboration, leading to a significant amount of data transmission due to redundant data. 
The amount of data transmission required for collaborative perception increases in a quadratic form relative to the number of participating CAVs. 
The edge-assisted paradigm employs a more efficient approach, where CAVs transmit perception data exclusively to the edge, which processes the aggregated data and distributes the consolidated results to all CAVs. 
Due to the assistance of the edge, all CAVs transmit perception data only once to complete collaboration.
As a result, the total amount of data transmission in the edge-assisted paradigm is substantially lower than that in the vehicle-based paradigm.
The LGCP framework further optimizes the efficiency through the proposed local-to-global management. 
Although multiple CAVs may be capable of perceiving an area, only a small subset transmits perception data related to the area.
Only leader CAVs are designated for data processing, which significantly reduces redundant transmission of perception data. 
In the deployments involving two CAVs, the LGCP framework achieves a notable transmission reduction of $45\times$, compared with the other two paradigms on average.
As the number of CAVs increases, the amount of data transmission under the three paradigms increases as well.
In the deployments involving four CAVs, the vehicle-based paradigm is overwhelmed with the excessive amount of data transmission. 
In a deployment involving seven CAVs, the total data transmission under the LGCP framework is only $8\times$ smaller than that of a complete CAV perception.
Compared to the edge-assisted paradigm, the LGCP framework achieves an average $44\times$ reduction.
% In deployments involving seven CAVs, the LGCP framework achieves a $49\times$ reduction, compared with the edge-assisted paradigm, with an amount of data transmission which is only $8\times$ smaller than that of a complete CAV perception.

\begin{figure}
    \centering
    \includegraphics[width=0.9\linewidth]{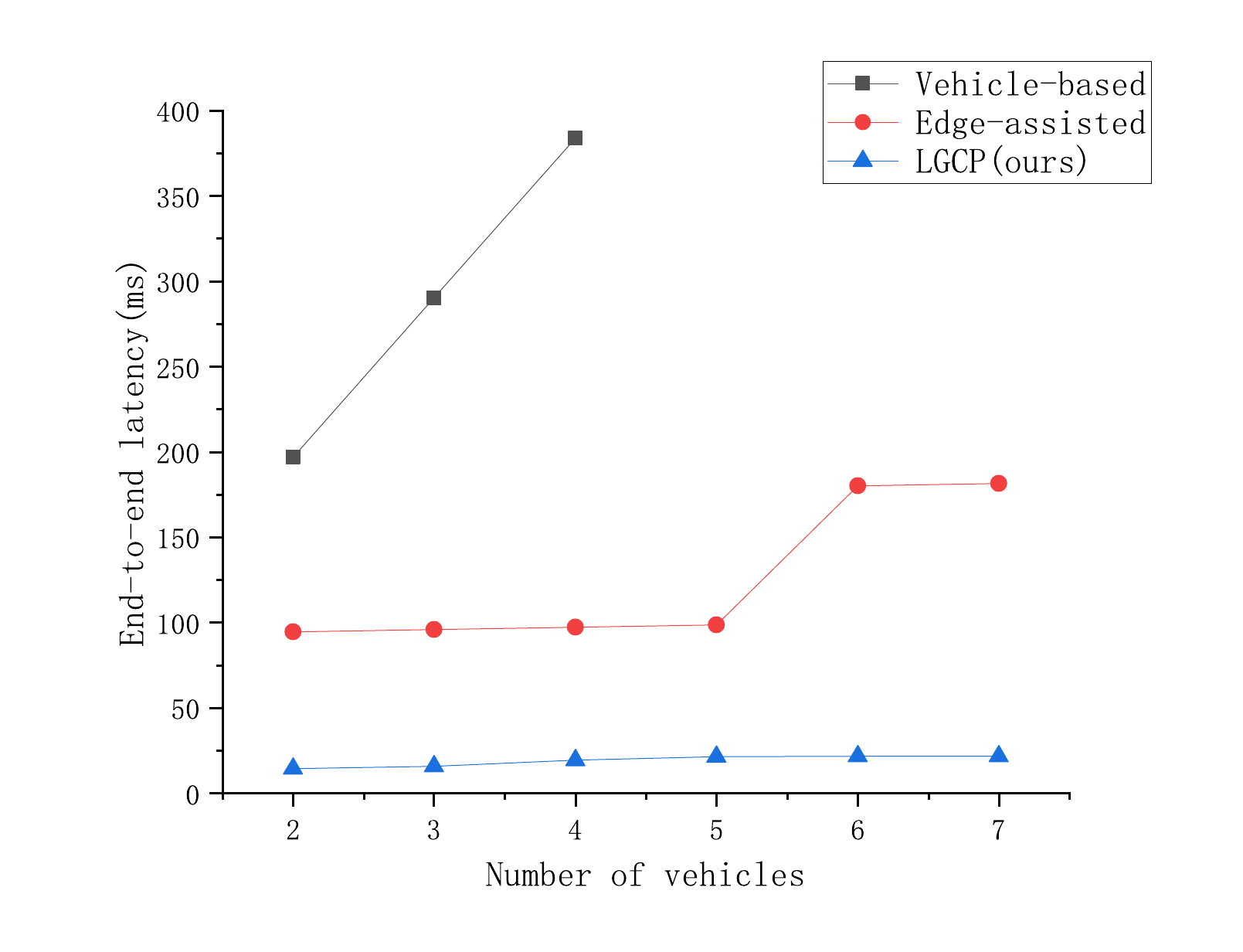}
    \caption{End-to-end latency for varying number of CAVs under the OPV2V dataset.}
    \label{fig:Latency1}
\end{figure}

Fig. \ref{fig:Latency1} shows the average end-to-end latency for all CAVs to complete collaboration for three different paradigms under the OPV2V dataset.
In the vehicle-based paradigm, an excessive amount of data transmission and a limitation of half-duplex make it impossible to complete the 10Hz real-time collaboration unless operating at a higher transmission rate.
Though the total amount of data transmission matches that of the edge-assisted paradigm when there are two CAVs, the end-to-end latency is approximately two times longer due to self-interference.
In the edge-assisted paradigm, the powerful edge computational capability and the exclusive upload setting enable support for collaboration among more CAVs. 
However, a linearly increasing fusion latency and a transmission bottleneck remain unavoidable as the number of CAVs increases.
The LGCP framework greatly reduces the amount of data transmission through scheduling, directly reducing the transmission latency.
It also decentralizes the computation process among CAVs, reducing the computation latency. 
% Since the number of collaboratively perceived areas remains constant, the edge server can select collaborators from more CAVs for these areas as the number of CAVs increases.
% Consequently, the latency only exhibits a slow increase rate, demonstrating excellent scalability performance.
When the number of collaborative CAVs reaches 4, our LGCP framework achieves a $384/20 = 19.2\times$ reduction in the end-to-end latency, compared with that of the vehicle-based paradigm and a $98/20=4.9\times$ reduction, compared with that of the edge-assisted paradigm.
When the number of collaborative CAVs reaches 7, our LGCP framework achieves approximately $182/22 \approx 8\times$ shorter latency, compared with that of the edge-assisted paradigm.

\begin{figure}
    \centering
    \includegraphics[width=0.9\linewidth]{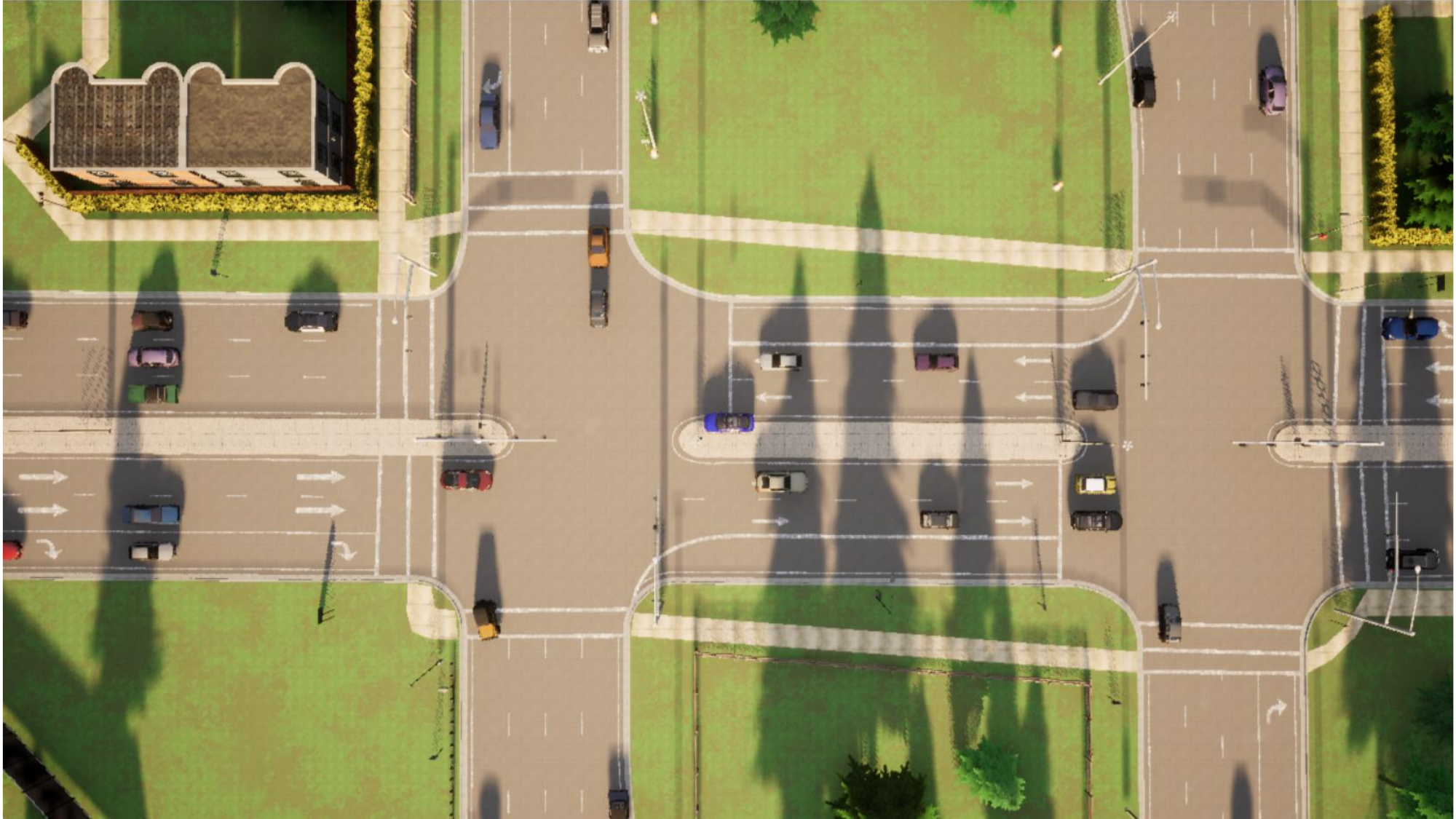}
    \caption{An overhead view from a CARLA simulation of traffic scenarios.}
    \label{fig:scene}
\end{figure}

Due to the limited number of CAVs in the OPV2V and V2XSet datasets, the evaluation of LGCP under multi-CAV conditions remains incomplete. 
To further investigate the end-to-end latency of LGCP in multi-CAV scenarios, we conduct a series of co-simulation experiments on the CARLA simulation, incorporating a number of CAVs from 5 to 30 and traffic dynamics. 
An example of the simulated urban environment used in our experiments is illustrated in Fig. \ref{fig:scene}. 

\begin{figure}
    \centering
    \includegraphics[width=0.9\linewidth]{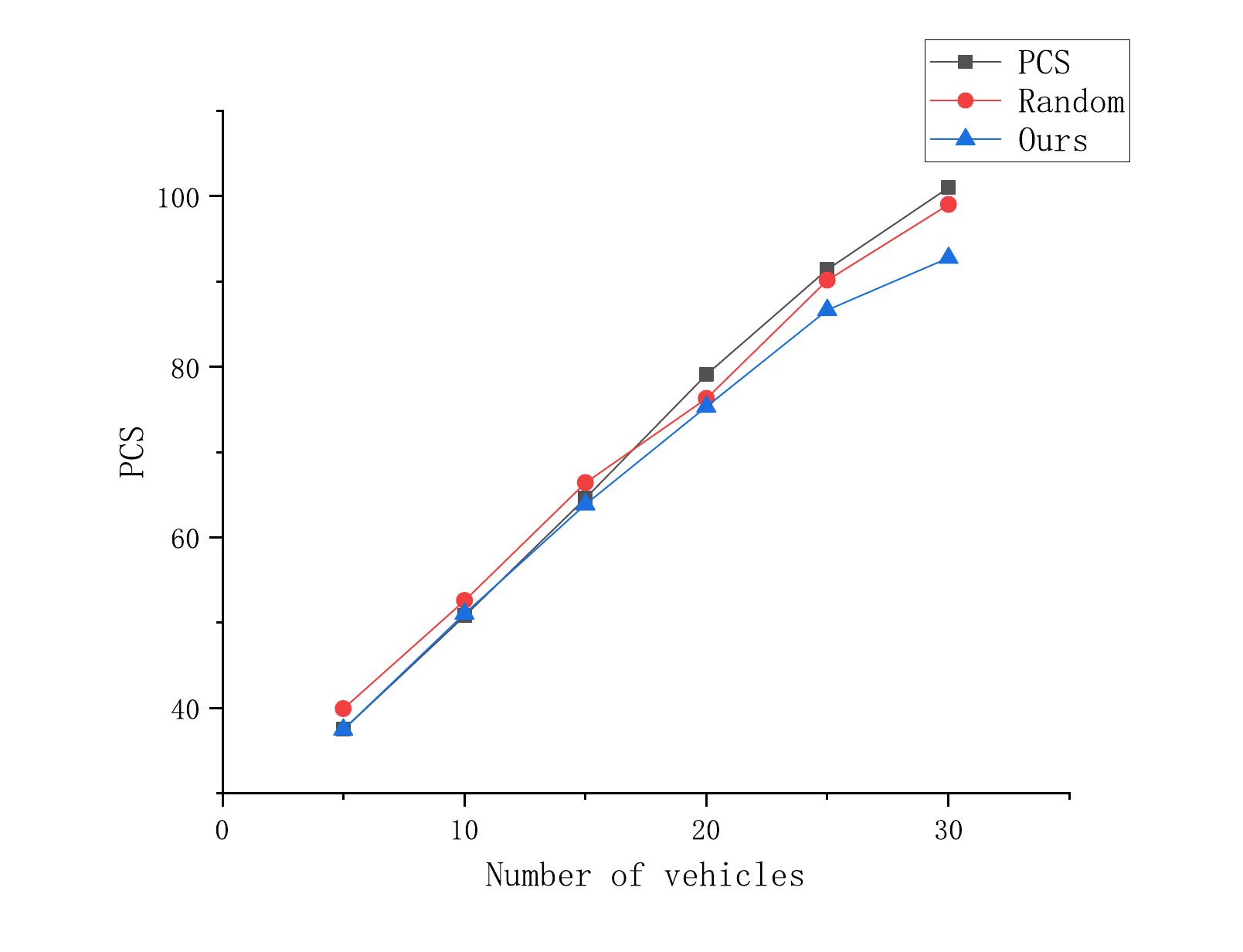}
    \caption{End-to-end latency for varying number of CAVs under co-simulation.}
    \label{fig:Latency}
\end{figure}

Fig. \ref{fig:Latency} shows the average end-to-end latency to complete collaboration in PCS, random, and our transmission scheduling algorithm under co-simulation.
The results show that our method is capable of completing collaborative perception within the required time budget in a scenario involving 30 CAVs.
In our scenario, where pairwise communication among CAVs may be needed, PCS suffers from excessive penalization of high-conflict links, especially in dense multi-CAV scenarios.
In contrast, our scheduling algorithm prioritizes transmission based on load, enabling more efficient utilization of limited resources and achieving higher efficiency.
Specifically, in the 30-CAV scenario, it reduces the end-to-end latency by 6.3\% and 8.1\% compared with random scheduling and PCS, respectively.
This demonstrates the effectiveness of our scheduling algorithm, especially in dense multi-CAV environments.

\section{conclusion}
In this paper, we proposed novel a local-to-global collaborative perception (LGCP) framework based on V2X communication and RSU assistance to enhance the perception efficiency in a communication- and computation-efficient manner. 
The RoI is partitioned into non-overlapping areas and a CAV group is assigned for collaboration around each area. 
Propagating the aggregated perception results of all areas to all CAVs, all CAVs can obtain a global view of the RoI.
The LGCP framework adopts a centralized scheduling policy, utilizes a RSU to assign CAV groups for areas, schedules the transmission of CAV groups, aggregates perception results of areas, and propagates the global result back to all CAVs.
Experimental results demonstrate that the proposed LGCP framework achieves an average 44× reduction in the amount of data transmission, while maintaining or even improving the overall collaborative performance.

\bibliographystyle{ieeetr}
\bibliography{LGCP}

\end{document}